\documentclass[showpacs,prl,twocolumn,amsmath,amssymb,superscriptaddress]{revtex4}

\usepackage{epsfig,amsmath}
\usepackage{subfigure}
\usepackage{graphicx}
\usepackage{dcolumn}
\usepackage{stmaryrd}
\usepackage{mathrsfs}
\usepackage{pifont}
\usepackage{amsthm}
\usepackage{amssymb}
\usepackage{bm}
\usepackage{latexsym}
\usepackage{hyperref}

\newcommand{\beq}{\begin{equation}}
\newcommand{\eeq}{\end{equation}}
\newcommand{\beqa}{\begin{eqnarray}}
\newcommand{\eeqa}{\end{eqnarray}}

\begin{document}

\title{Quantum states experimentally achieving high-fidelity transmission over a spin chain}

\author{Zhao-Ming Wang}
\affiliation{Department of Physics, Ocean University of China, Qingdao, 266100, China}
\affiliation{Department of Theoretical Physics and History of Science, The Basque
Country University(EHU/UPV) and IKERBASQUE, Basque Foundation for Science,
48011 Bilbao, Spain}

\author{Lian-Ao Wu}
\affiliation{Department of Theoretical Physics and History of Science, The Basque
Country University(EHU/UPV) and IKERBASQUE, Basque Foundation for Science,
48011 Bilbao, Spain}

\author{C. Allen Bishop}

\thanks{This submission was written by the author acting in his own
 independent capacity and not on behalf of UT-Battelle, LLC, or its affiliates or successors.}

\affiliation{Center for Quantum Information Science, Computational Sciences and Engineering Division,
Oak Ridge National Laboratory, Oak Ridge, Tennessee 37831-6418, USA}

\author{Yong-Jian Gu}

\affiliation{Department of Physics, Ocean University of China, Qingdao, 266100, China}

\author{Mark S. Byrd}
\affiliation{Department of Physics, Southern Illinois University, Carbondale,
Illinois 62901-4401, USA}


\begin{abstract}
A uniformly coupled double quantum Hamiltonian for a spin chain has recently been implemented experimentally.  We propose
a method for the determination of initial quantum
states that will provide perfect or near-perfect state transmission
for an arbitrary Hamiltonian including this one.  By calculating the eigenvalues and eigenvectors of a unitary operator obtained from the free evolution plus
an exchange operator, we find that the double quantum Hamiltonian spin chain will support a three-spin initial encoding that
will transfer along the chain with remarkably high fidelity.
The fidelity is also found to decrease very slowly with increasing
chain length.  In addition, we are able to explain previous results showing  exceptional transfer using this method.
\end{abstract}


\maketitle

{\em Introduction.---}
Quantum information processing (QIP) often requires the transfer of known or unknown
quantum states from one subspace to another within an information processing device. In recent years, the quantum
spin chain has become a prime candidate for quantum communication purposes
such as these \cite{Bose2003,Wu20091,Yung2003}. In the simplest configuration,
where the nearest neighbor couplings are considered to be equal, perfect
state transmission is typically not possible between two single spin processors within a
linear chain. In other words, there is typically
a non-vanishing probability that the initial excitation amplitude can be
found outside the receiving spin location \cite{Wiesniak} at any given time. In principle, however, perfect state transfer (PST) can be
realized by properly engineering the couplings between neighboring sites
\cite{Christandl2004}. High fidelity state transmissions can also be obtained
using weakly coupled external qubits \cite{Wojcik05,Oh2011}, modifying only
 one or two couplings \cite{Apollaro2012,Banchi2011}, or by encoding
the states using multiple spins \cite{Osborne2004,Haselgrove,
Allcock2009,Wang2009,Allen2010}.  In
\cite{Wang2009,Allen2010}, a class of states were found
to transfer very well across long XY coupled spin chains.  The
existence of PST has also been established for a variety of interacting
media, including, but not limited to, the spin chain model \cite{Wu20092}.
Recently, exact state swap through a spin ring has
been investigated. It was shown that there is a straightforward approach
to calculating the probability of the occurrence of an exact state swap
\cite{Liu2012}.

The schemes developed in Ref. \cite{Wu20092} prompted the following
question. Given an arbitrary spin chain Hamiltonian, can we find initial
states which can be used to enable high-fidelity state transmission?
In this letter, we answer this question and show that for a uniformly
coupled chain, there exists a particular state which reliably transfers quantum
information over large distances. We use a multi-spin encoding scheme
and find the existence of a three-spin encoding which can provide reliable state transmission. In this case, the
encoding and decoding processes can also be realized easily
\cite{Wang2009}. This report is therefore important from an experimental perspective
due to the ease of implementation which is typically favorable.

{\em The method for identifying high-fidelity states.---}
Consider a spin chain consisting of $N$ sites which evolves
according to some Hamiltonian $H$ in a single
excitation subspace. Suppose for the moment the initial state of our
system is $\left\vert \Psi (0)\right\rangle =\left\vert \mathbf{1}\right\rangle
=\left\vert 1\right\rangle _{A}\otimes \left\vert 101...1\right\rangle
\otimes \left\vert 0\right\rangle _{B}$, where A and B denote
separate processors. After the system evolves, the state at time $t$ will be
\begin{equation}
\left\vert \Psi (t)\right\rangle =U(t)\left\vert \mathbf{1}\right\rangle
=\exp (-iHt)\left\vert \Psi (0)\right\rangle,
\end{equation}%
where $\hbar $ is taken to be 1 throughout. Suppose that at some time
$\tau $ PST occurs, then
\begin{equation}
\left\vert \Psi (\tau )\right\rangle =U(\tau )\left\vert \mathbf{1}%
\right\rangle =\left\vert \mathbf{N}\right\rangle,  \label{Eq:1}
\end{equation}
where $\left\vert \mathbf{N}\right\rangle=\left\vert 0\right\rangle _{A}\otimes \left\vert 101...1\right\rangle
\otimes \left\vert 1\right\rangle _{B}$.
We can use a permutation operator $P_{AB}$ to swap all states in A and B,
then the quantum information can be transferred from A to B. The permutation
operator can be expressed as:%
\begin{equation}
P_{AB}=\sum\nolimits_{\alpha \beta }(\left\vert \beta _{A}\right\rangle
\left\langle \alpha _{A}\right\vert \otimes \left\vert \alpha
_{B}\right\rangle \left\langle \beta _{B}\right\vert ),
\end{equation}%
where $\alpha ,\beta =1,2,...,2^{k}$ represent the standard basis for the $k$ qubits located in processors A
and B. Clearly $P_{AB}^{\dagger} = P_{AB}$ and $P_{AB}^{2}=1$. $\left\vert \alpha (\beta )_{A(B)}\right\rangle $
refers to a state $\left\vert \alpha (\beta )\right\rangle $ in processor
A (B). From Eq. (\ref{Eq:1})
\begin{equation}
U(\tau )\left\vert \mathbf{1}\right\rangle =P_{AB}\left\vert \mathbf{1}%
\right\rangle,
\end{equation}%
Then
\begin{equation}
P_{AB}U(\tau )\left\vert \mathbf{1}\right\rangle =W(\tau )\left\vert \mathbf{%
1}\right\rangle =\left\vert \mathbf{1}\right\rangle.  \label{Eq:6}
\end{equation}

We introduce the unitary operator $W(\tau )=P_{AB}U(\tau )$. From Eq. (\ref%
{Eq:6}), if the state $\left\vert \mathbf{1}\right\rangle $ is an
eigenvector of the operator $W$ at time $\tau $, PST occurs. The
eigenvectors of $W$ reveal information about the possibilities of a
specific state transmission. The problem of solving Schr\"{o}dinger's equation
now becomes a standard eigen-problem of the operator $W$.

Since $W(\tau )$ is a unitary operator it has a complete set of
orthonormal eigenvectors $ \{\left\vert \Psi _{m}(0)\right\rangle \}_{\tau }$
corresponding to eigenvalues $\{E_{m}\}_{\tau }$,
\begin{equation}
W(\tau )\left\vert \Psi _{m}(0)\right\rangle =E_{m}\left\vert \Psi
_{m}(0)\right\rangle.
\end{equation}
This can also be written as
\begin{equation}
U(\tau )\left\vert \Psi _{m}(0)\right\rangle =E_{m}P^{\dag }_{AB}\left\vert \Psi
_{m}(0)\right\rangle,
\end{equation}
where $U(\tau )\left\vert \Psi _{m}(0)\right\rangle $ is the wave function $%
\left\vert \Psi _{m}(\tau )\right\rangle $ of the system which was initially prepared
in the eigenstate $\left\vert \Psi _{m}(0)\right\rangle $.
If $\left\vert \Psi _{m}(0)\right\rangle $ is a product state
\begin{equation}
\left\vert \Psi _{m}(0)\right\rangle =\left\vert A\right\rangle \otimes
\left\vert C\right\rangle,  \label{Eq:product}
\end{equation}
with $\left\vert A\right\rangle $ describing the state of processor A
 and $\left\vert C\right\rangle $ describing the rest of the system, we can then obtain
\begin{equation}
\left\vert \Psi _{m}(\tau )\right\rangle =E_{m}P^{\dag }_{AB}\left\vert
A\right\rangle \otimes \left\vert C\right\rangle
=E_{m}\left\vert B\right\rangle \otimes \left\vert C^{\prime}\right\rangle.
\end{equation}

 For the single
excitation subspace, if one of the eigenvectors $\left\vert \Psi
_{m}(0)\right\rangle =\left\vert \mathbf{1}\right\rangle $ at time $\tau $,
PST occurs. If the eigenvectors are degenerate, an arbitrary
linear superposition of these degenerate states is also suitable for PST. Suppose
there are $L$ degenerate eigenvectors $\left\vert \Psi _{l}(0)\right\rangle
(l=1,2,...L)$, which have common eigenvalues $E_{L}$. The state

\begin{equation}
\left\vert \Psi (0)\right\rangle =\sum\limits_{l=1}^{L}C_{l}\left\vert \Psi
_{l}(0)\right\rangle.
\end{equation}%
is an eigenvector of $W(\tau )$, where $C_{l}$ is an arbitrary number.
Our analysis describes a method for finding a state which can
realize PST. (Note that these states are not all unique.)  For a given Hamiltonian, if we initially prepare the state
$\left\vert \Psi (0)\right\rangle $ as an eigenvector of the operator $W(\tau )$,
then after time $\tau$ PST occurs.

\begin{figure}[tbph]
\centering
\includegraphics[scale=0.4,angle=0]{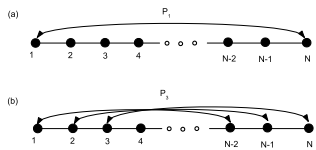}
\caption{Schematic of our quantum transmission protocol: (a) single-site encoding  (b) three-site encoding.}
\label{fig:1}
\end{figure}

For state transmission from one end spin $1$ to another end spin $N$, the
exchange operator is given by $P_{1}=\left\vert 1\right\rangle \left\langle
N\right\vert +\left\vert N\right\rangle \left\langle 1\right\vert +P_{0}$
which is shown in Fig.~\ref{fig:1},
where $P_{0}=\sum_{j}\left\vert j\right\rangle \left\langle j\right\vert $($j\neq1,N$).  We now considered a particular Hamiltonian, but emphasize that our method can be used for any
Hamiltonian not only this example.

{\it An experimentally implementable Hamiltonian.}--
Now consider the recently implemented Hamiltonian
called a double quantum (DQ) Hamiltonian \cite{Ramanathan}:
\begin{equation}
H=-\sum_{i=1}^{N-1}J_{i,i+1}(X_{i}X_{i+1}-Y_{i}Y_{i+1}).\label{Eq:q}
\end{equation}
where $J_{i,i+1}$ denotes the coupling between
sites $i$ and $i+1$.
This nearest neighbor coupled one-dimensional spin chain can be
experimentally implemented using solid-state nuclear magnetic resonance \cite%
{Ramanathan,Fiori,Zhang} in $^{19}$F spins in a crystal of fluorapatite ((FAp-Ca$_{5}$%
(POa)$_{3}$)F) \cite{Fiori,Zhang}.
The system described by Eq.~(\ref{Eq:q}) will exhibit free evolution such that
the evolution operator at time $\tau $ will be $U(\tau )=$exp$[-i\tau H]$.
We can diagonalize the Hamiltonian $H$ such that $H_{d}=W^{\dag }HW$ in
the single
excitation subspace. The evolution operator can therefore be expressed by $%
U(\tau )=W$exp$[-i\tau H_{d}]W$ and the $N$ eigenvectors of $W(\tau )$ can be obtained as a function of $\tau $.
Furthermore, we consider a natural configuration for a DQ
Hamiltonian with open ends. The $z$%
-component of the total for the staggered spins is a conserved quantity, $[(\sum_{i\in
odd}Z_{i}-\sum_{i\in even}Z_{i}),H]=0$. For simplicity, we
will only
consider the single excitation subspace of the full Hilbert space.
In this case the
total number of flipped spins is one.  The basis for this
subspace will be denoted as $\left\vert \mathbf{j}%
\right\rangle$ which indicates that, after flipping, the even (odd) site spins
 all of the spins reside
in the $\left\vert 0\right\rangle$ ($\left\vert 1\right\rangle$) state except for
the spin at site $j$ which is in the $\left\vert 1\right\rangle$ ( $\left\vert
0\right\rangle$)state.
For example, in a $N=5$ site chain, the single excitation
subspace will be spanned by $\left\vert \mathbf{1}%
\right\rangle$=$\left\vert 11010\right\rangle
,\left\vert \mathbf{2}%
\right\rangle=\left\vert 00010\right\rangle$,..., etc. If we flip the
even numbered states we find that the total up spin is
actually one. We will use this description throughout this paper.

{\em Example I: nonuniform couplings---} We will consider several different coupling configurations with the potential for
high-fidelity state transmission and the best results will be provided at the
end of our analysis. First as an example, we consider two pre-engineered couplings:
(1) weak couplings at both ends, where $J_{1,2}=J_{N-1,N}=J_{0}$ and $J_{i,i+1}=J$
elsewhere. (2) couplings termed PST, where $J_{i,i+1}=\sqrt{i(N-i)}$.
It is already known that high fidelity ($F_{max}\approx1$) state transmission
 for the first configuration
\cite{Wojcik05, Oh2011}
and perfect fidelity ($F_{max}=1$) for the second configuration can
be gained in a spin system \cite{Christandl2004}. Here we will use these two
kinds couplings to show the applicability of our methods.

\begin{table}[htbp]
\centering \doublerulesep 0.5pt
\begin{tabular}{ccccccc}
\hline
&  & $\left\vert \mathbf{1}\right\rangle$ & $\left\vert \mathbf{2}%
\right\rangle$ & $\left\vert \mathbf{3}\right\rangle$ & $\left\vert \mathbf{4%
}\right\rangle$ & $\left\vert \mathbf{5}\right\rangle$ \\ \hline
1 & (0.999,0.03) & 0.035 & -0.500 & 0.705 & -0.500 & 0.035 \\ \hline
2 & (0.999,-0.03) & 0.035 & 0.500 & 0.705 & 0.500 & 0.035 \\ \hline
3 & (1.000,0) & 0.398 & 0 & -0.066 & 0 & 0.914 \\ \hline
4 & (1.000,0) & 0.999 & 0 & -0.050 & 0 & -0.003 \\ \hline
5 & (-1.000,0) & 0 & 0.707 & 0 & -0.707 & 0 \\ \hline
\end{tabular}%
\caption{The complex eigenvalues and corresponding eigenvectors of the operator $%
W(\protect\tau)$ at $\protect\tau=31$ in a spin chain where the weak coupling
conditions $J_{1,2}=J_{4,5}=0.1J$ are satisfied. We take $N$=5, $J=-1$.}
\label{t1}
\end{table}

For a five-spin system with weak couplings at both ends, we take $%
J_{1,2}=J_{4,5}=0.1J$. $J$ equals -1 elsewhere. The eigenvalues and eigenvectors of the operator $%
W(\tau )$ at an arbitrary time $\tau $ can be obtained numerically.
We will consider those which span the single-excitation subspace.
In Table~\ref{t1}, we plot the results for $\tau =31$. The first
column labels the complex eigenvalues while the remaining
columns are associated with the amplitudes of the states at the top of
each column.
The coefficients
in columns 2-6 could be complex numbers but our
results show that the imaginary part always equals zero, so we take them to be
real throughout. The same meaning holds for Tables \ref{t1}, \ref{t2}, %
\ref{t4}, and \ref{t5}.  For example, at the line labeled with a 4 the eigenvalue
is (1.000,0) and the eigenvector is $0.999\left\vert \mathbf{1}\right\rangle
-0.050\left\vert \mathbf{3}\right\rangle -0.003\left\vert \mathbf{5}%
\right\rangle $. The state $\left\vert \mathbf{1}\right\rangle$ closely approximates this eigenvector
and it can be written as the aforementioned product state. If we use the
state $\left\vert \mathbf{1}\right\rangle $ as the initial state of the
whole system, then at time $\tau =31$, the system will closely approximately
the state $\left\vert
\mathbf{N}\right\rangle $.

\begin{figure}[htbp]
\centering
\includegraphics[scale=0.36,angle=0]{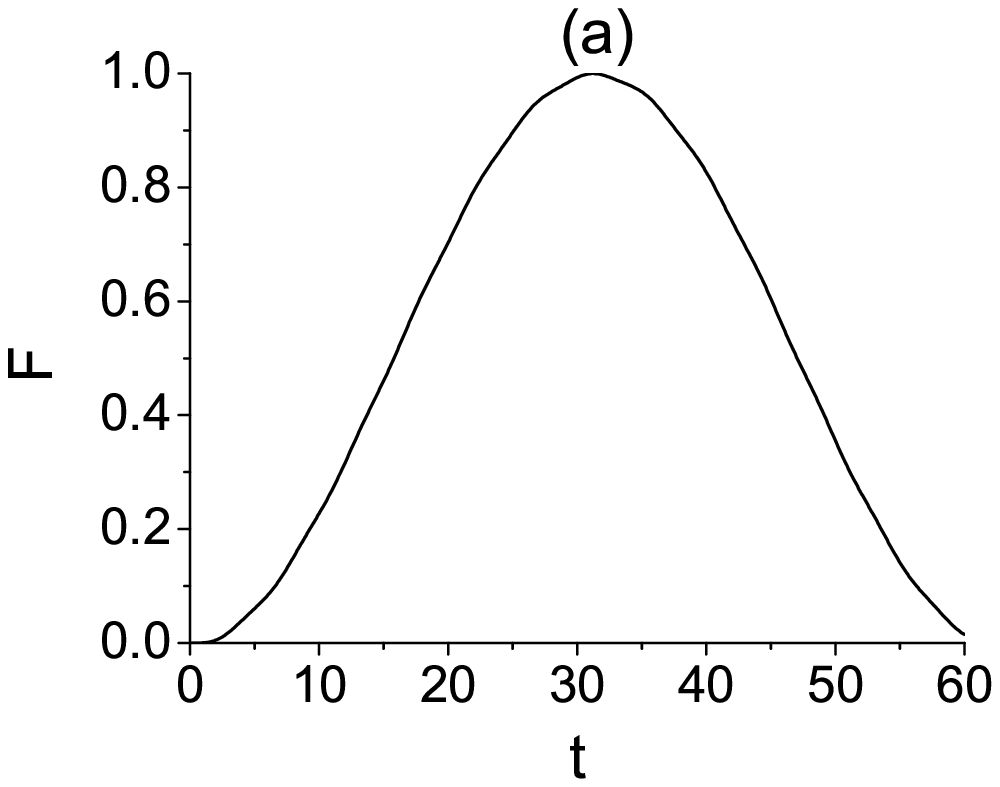} %
\includegraphics[scale=0.37,angle=0]{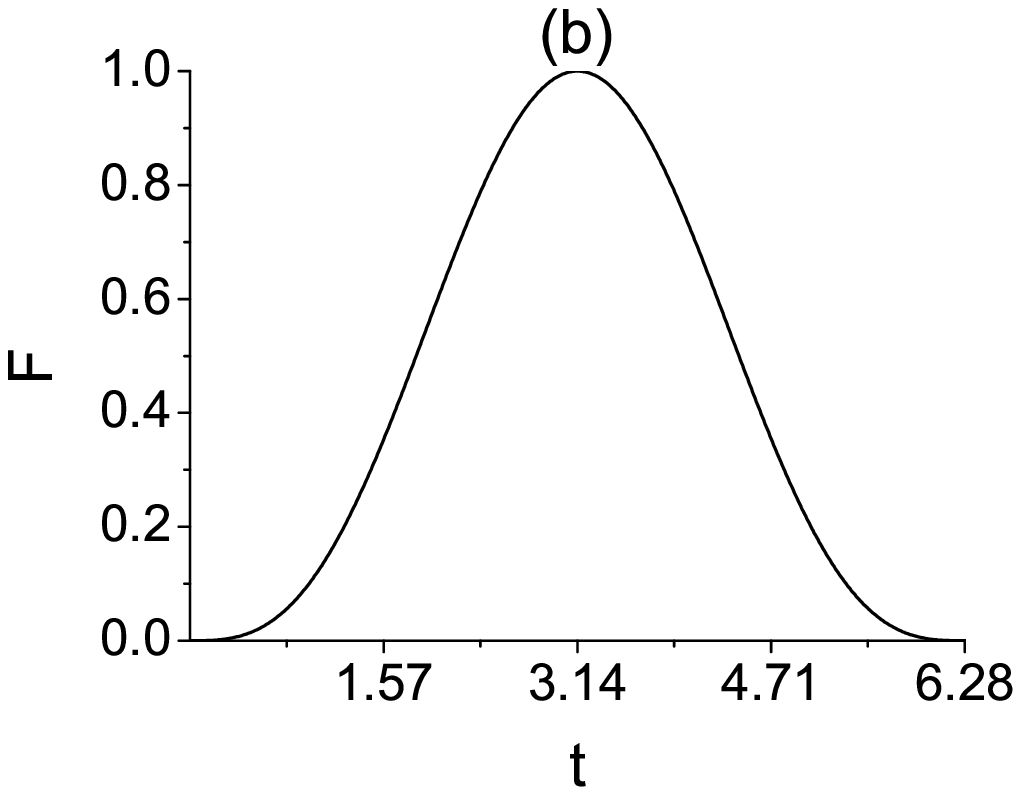}
\caption{The fidelity as a function of time for (a) $N=5$, weak coupling
conditions for $J_{1,2}=J_{4,5}=0.1J$ (b) $N=4$, channel coupling
conditions $J_{i,i+1}=\sqrt{i(N-i)}$. See text for more details.}
\label{fig:2}
\end{figure}

We use the fidelity between the received state and the ideally transfered state,
$F=\sqrt{\left\langle \Phi (0)\right\vert \rho (t)\left\vert
\Phi (0)\right\rangle }$ as a measure of the quality of the transfer. Here $\left\vert \Phi (0)\right\rangle $ is a
state at the receiving end which has the same form as the state initially prepared by the sender. $\rho (t)$ is
the reduced density matrix of the receiver's spin at time $t$ and is obtained by
tracing over all but the receiver's sites. In Fig. \ref{fig:2} (a) we plot
the fidelity versus time $t$ for the weakly coupled chain.
The initial state is $\left\vert \mathbf{1}\right\rangle $ has the maximum fidelity, $F=1$ at time $t=31$.

\begin{table}[htbp]
\centering \doublerulesep 0.5pt
\begin{tabular}{cccccc}
\hline
&  & $\left\vert \mathbf{1}\right\rangle$ & $\left\vert \mathbf{2}%
\right\rangle$ & $\left\vert \mathbf{3}\right\rangle$ & $\left\vert \mathbf{4%
}\right\rangle$ \\ \hline
1 & (0,1.000) & 0.707 & 0 & 0 & -0.707 \\ \hline
2 & (0.018,-1.000) & 0 & 0.707 & -0.707 & 0 \\ \hline
3 & (-0.009,1.000) & 0.612 & -0.354 & -0.354 & 0.612 \\ \hline
4 & (0.028,1.000) & 0.354 & 0.612 & 0.612 & 0.354 \\ \hline
\end{tabular}%
\caption{The complex eigenvalues and corresponding eigenvectors of the operator $%
W(\protect\tau)$ at $\protect\tau=3.14$ in a $N$=4 site spin chain
with couplings given by $J_{i,i+1}=\sqrt{i(N-i)}$.}
\label{t2}
\end{table}

 Next we discuss PST. For the simple case of $N=4$, the results at time $\tau =3.14$ are listed
in Table \ref{t2}. None of the eigenvectors can be written in the form of a
product state $\left\vert \Psi _{m}(0)\right\rangle =\left\vert
A\right\rangle \otimes \left\vert C\right\rangle $, but the eigenvalues of
1, 3, 4 are roughly degenerate. Consider the superposition
\begin{equation}
\sqrt{2}\left\vert \Psi _{1}(0)\right\rangle +\sqrt{\frac{3}{8}}\left\vert
\Psi _{3}(0)\right\rangle +\sqrt{\frac{1}{8}}\left\vert \Psi
_{4}(0)\right\rangle =\left\vert \mathbf{1}\right\rangle,
\end{equation}
where
\begin{eqnarray}
\left\vert \Psi _{1}(0)\right\rangle &=&\sqrt{\frac{1}{2}}(\left\vert
\mathbf{1}\right\rangle -\left\vert \mathbf{4}\right\rangle ),  \notag \\
\left\vert \Psi _{3}(0)\right\rangle &=&\sqrt{\frac{3}{8}}(\left\vert
\mathbf{1}\right\rangle +\left\vert \mathbf{4}\right\rangle )-\sqrt{\frac{1}{%
8}}(\left\vert \mathbf{2}\right\rangle +\left\vert \mathbf{3}\right\rangle
),
\\
\left\vert \Psi _{4}(0)\right\rangle &=&\sqrt{\frac{1}{8}}(\left\vert
\mathbf{1}\right\rangle +\left\vert \mathbf{4}\right\rangle )+\sqrt{\frac{3}{%
8}}(\left\vert \mathbf{2}\right\rangle +\left\vert \mathbf{3}\right\rangle
).
\notag
\end{eqnarray}

The state $\left\vert 1\right\rangle $ at site 1 can be transferred exactly
to site 4 at time $\tau =3.14$. In Fig. \ref{fig:2}(b) we plot the time
evolution of the fidelity when transferring a state $\left\vert
1\right\rangle $ from site 1 to 4. We also see that at time $\tau =3.14$
the fidelity is nearly 1. These examples illustrate the validity and practicality of our method while providing a general method to obtain the results.

{\em Example II: uniform couplings. ---}Consider the most
natural configuration for a spin chain; a uniformly coupled spin chain. We take the ferromagnetic coupling $J_{i,i+1}=J=-1$.  Note that PST is typically
unattainable in these systems using single-spin encodings \cite%
{Christandl2004,Osborne2004}. We will consider both single spin
encodings as well as multi-spin encodings. For single-spin encodings, our
calculations confirm that PST cannot occur in this model as is known
\cite{Liu2012}.
In table \ref{t3} we list the the maximal $p_{m}$
and the corresponding $\tau $ for a $N=7$ uniform chain for different eigenvectors.
Here $%
p_{m}$ is defined as the overlap between the eigenvector $\left\vert \Psi
_{m}(0)\right\rangle $ and the initial state $\left\vert \mathbf{1}%
\right\rangle $, i.e., $p_{m}=\left\vert \left\langle \Psi _{m}(0)\right\vert
\mathbf{1}\rangle\right\vert $.

\begin{table}[htbp]
\centering \doublerulesep 0.5pt
\begin{tabular}{cccccccc}
\hline
$\left\vert \Psi _{m}\right\rangle$ & $\left\vert \Psi _{1}\right\rangle$ & $%
\left\vert \Psi _{2}\right\rangle$ & $\left\vert \Psi _{3}\right\rangle$ & $%
\left\vert \Psi _{4}\right\rangle$ & $\left\vert \Psi _{5}\right\rangle$ & $%
\left\vert \Psi _{6}\right\rangle$ & $\left\vert \Psi _{7}\right\rangle$ \\
\hline
$\tau$ & 31.1 & 8 & 31 & 18.6 & 35.6 & 30.6 & 8.9 \\ \hline
$p_{m}$ & 0.7071 & 0.6295 & 0.7062 & 0.6402 & 0.7068 & 0.6842 & 0.7071 \\
\hline
\end{tabular}%
\caption{ The maximal $p_{m}$ and corresponding values of $\protect\tau$. $N$%
=7, the maximum values are found in a time interval $[5,40]$.}
\label{t3}
\end{table}

Now we will examine multi-spin encoding schemes. As an example, we first consider a
three-spin encoding. Specifically, suppose we wish to transfer a state of the form $\left\vert \Psi (0)\right\rangle
=(\alpha \left\vert 100\right\rangle +\beta \left\vert 010\right\rangle
+\gamma \left\vert 001\right\rangle )_{A}\otimes \left\vert
00...0\right\rangle \otimes \left\vert 000\right\rangle _{B}$. As shown in Fig. \ref{fig:1}(b), we intend to
transfer the state of the first three spins to the opposite end. The results for a $N=6$ site chain
are given in Table \ref{t4} for time $\tau =4.0$. The eigenvalues of 1 and 2
are roughly degenerate and the approximate relation
\begin{equation}
\left\vert \Psi _{1}(0)\right\rangle +\left\vert \Psi
_{2}(0)\right\rangle=-\left\vert \mathbf{1}\right\rangle +\left\vert \mathbf{%
3}\right\rangle
\end{equation}
can be written in the form of a product state $(-\left\vert 110\right\rangle
+\left\vert 011\right\rangle )_{A}\otimes \left\vert 10...1\right\rangle $.
The state $(-\left\vert 110\right\rangle +\left\vert 011\right\rangle )/%
\sqrt{2}$ is therefore suitable for transmission. We have also checked the case where $N=7$ and
find that at time $\tau =28.8$ the above states can be obtained again.

\begin{table}[htpb]
\centering \doublerulesep 0.5pt
\begin{tabular}{cccccccc}
\hline
&  & $\left\vert \mathbf{1}\right\rangle$ & $\left\vert \mathbf{2}%
\right\rangle$ & $\left\vert \mathbf{3}\right\rangle$ & $\left\vert \mathbf{4%
}\right\rangle$ & $\left\vert \mathbf{5}\right\rangle$ & $\left\vert \mathbf{%
6}\right\rangle$ \\ \hline
1 & (0.117,-0.993) & -0.493 & 0.005 & 0.500 & 0.500 & 0.005 & -0.500 \\
\hline
2 & (-0.117,-0.993) & -0.493 & -0.005 & 0.500 & -0.500 & 0.005 & 0.500 \\
\hline
3 & (0.252,0.968) & -0.275 & 0.590 & -0.275 & -0.275 & 0.590 & -0.275 \\
\hline
4 & (-0.252,0.968) & 0.275 & 0.590 & 0.275 & -0.275 & -0.590 & -0.275 \\
\hline
5 & (0.544,0.839) & 0.421 & 0.379 & 0.405 & 0.405 & 0.379 & 0.421 \\ \hline
6 & (-0.54,0.839) & -0.421 & 0.379 & -0.405 & 0.405 & -0.379 & 0.421 \\
\hline
\end{tabular}%
\caption{The eigenvalues and corresponding eigenvectors of the operator $%
W(\protect\tau)$ at $\protect\tau=4.0$ using a 3 spin encoding. Here $N$%
=6.}
\label{t4}
\end{table}

\begin{table}[htbp]
\centering \doublerulesep 0.5pt
\begin{tabular}{ccccccc}
\hline
&  & $\left\vert \mathbf{1}\right\rangle$ & $\left\vert \mathbf{2}%
\right\rangle$ & $\left\vert \mathbf{3}\right\rangle$ & $\left\vert \mathbf{4%
}\right\rangle$ & $\left\vert \mathbf{5}\right\rangle$ \\ \hline
1 & (0.999,-0.025) & -0.263 & -0.263 & 0.850 & -0.263 & -0.263 \\ \hline
2 & (-0.999,0.041) & 0.500 & -0.500 & 0.000 & -0.500 & 0.500 \\ \hline
3 & (-0.997,0.076) & 0.500 & -0.500 & 0.000 & 0.500 & -0.500 \\ \hline
4 & (0.997,0.076) & 0.500 & 0.500 & 0.000 & -0.500 & -0.500 \\ \hline
5 & (0.998,0.067) & 0.425 & 0.425 & 0.526 & 0.425 & 0.425 \\ \hline
\end{tabular}%
\caption{The eigenvalues and corresponding eigenvectors of the operator $%
W(\protect\tau)$ at $\protect\tau=47.2$ using a 2-spin encoding. Here $N$%
=5. }
\label{t5}
\end{table}


\begin{figure}[htbp]
\centering
\includegraphics[scale=0.4,angle=0]{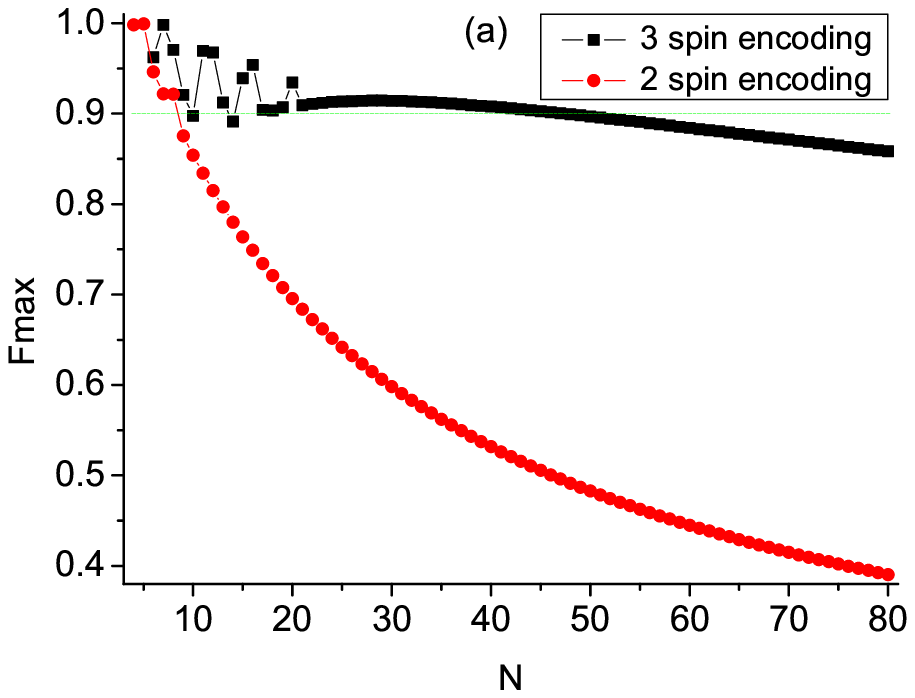} %
\includegraphics[scale=0.4,angle=0]{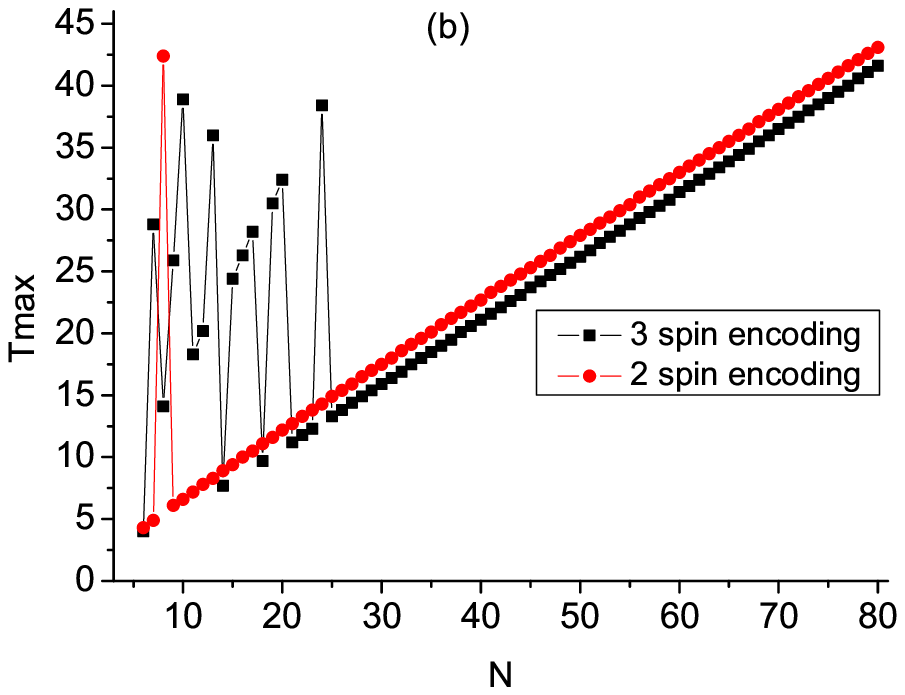}
\caption{(Color online.) Length dependence of the maximum fidelity
achievable $F_{max}$ and the associated arrival times $T_{max}$ for the state (a) $%
(-\left\vert 110\right\rangle +\left\vert 011\right\rangle )/\protect\sqrt{2}
$ and (b) $(\left\vert 11\right\rangle -\left\vert00\right\rangle )/\protect\sqrt{2%
}$. The time is searched within the interval $[0,50]$.}
\label{fig:3}
\end{figure}

Table \ref{t5} lists the results corresponding to a $N=5$ chain for the
two-spin encoding.
At $\tau =47.2$, the relation $\left\vert \Psi _{2}(0)\right\rangle
+\left\vert \Psi _{3}(0)\right\rangle=\left\vert \mathbf{2}\right\rangle
-\left\vert \mathbf{1}\right\rangle$ approximately holds which can also be written in the form of the product
state in Eq.~(\ref{Eq:product}).

We have found some states realizing high-fidelity state transmission,
for small $N$.  Now we will check to see if the state $(-\left\vert 11\right\rangle +\left\vert
00\right\rangle )/\sqrt{2}$ and $(-\left\vert 110\right\rangle +\left\vert
011\right\rangle )/\sqrt{2}$ can be transferred with high
fidelity across chains of arbitrary length $N$. In Fig.~\ref{fig:3}, we plot the
maximum fidelity $F_{max}$ and the associated arrival time $T_{max}$ as a function of chain length $N$. The
analytic expression with eigenvalues $E_{m}=-2J\cos [\pi m/(N+1)]$ and
eigenvectors $\left\vert \Psi _{m}(0)\right\rangle =\sqrt{2/(N+1)}%
\sum\nolimits_{j}\sin (q_{m}j)\left\vert \mathbf{j}\right\rangle $ are used.
For practical implementation of our protocol, the maximum fidelity
is found in the time [0,50]. For the two-spin encoding, the high fidelity
associated with short chain lengths cannot be acheived with increasing chain
length. $F_{max}$ quickly decreases with increasing $N$. However,
this robustness can be observed even for long chains using the
three-spin encoding. The fidelity is exceptionally large for a
relatively long chain. Therefore, using this state, a high-fidelity
state transfer can be gained. $F_{max}=0.96$ for $N=6$ at $t=4.0$, $F_{max}=1.00$ for $N=7$ at $t=28.8$ which agrees
with our previous analysis \cite{Wang2009,Allen2010}. Note that we only consider two and three-spin encodings here.
For encodings using more than three spins, we conjecture that for odd spin encodings
some states can be found to possess high-fidelity transmission even over long chains. From Fig. \ref{fig:3}~(b) we find that the arrival
time $T_{max}$ typically increases with increasing chain length $N$ except for some deviation with small values of $N$. We also find that the $T_{max}$ associated with the
three-spin encoding is a little longer than in the two-spin encoding
case for $N>24$. This suggests that encodings using larger Hilbert spaces require longer
waiting times for the maximum fidelity.

{\em Conclusions.--}In conclusion, we have introduced a method to find  states which can be
transmitted through spin channels with high fidelity. The method can be easily
implemented numerically and can be applied
to $N$-site encodings, with $N$ arbitrary.
Using our method we have provided examples for the DQ Hamiltonian which exhibit uniform and nonuniform exchange couplings.
For the uniform chain, a 3-spin encoding
$(-\left\vert 110\right\rangle +\left\vert 011\right\rangle )/\sqrt{2}$
was found to exhibit high fidelity state transmission. Using a
simple similarity transformation \cite{Ramanathan}, our results can be extended to
the standard Heisenberg XY model. In this case we have provided an explanation for the appearence of the class of
initial states which were previously discovered \cite{Wang2009,Allen2010}.  These states are exceptional due to the fact that they use simple encodings and transfer extremely well.  Our work therefore provides a new method, new results, and an explanation of previously known important results.

\begin{acknowledgements}
This work is supported by NSFC (Grant No.
11005099), Fundamental Research Funds for the Central
Universities (Grant No. 201013037), an Ikerbasque Foundation
Startup, the Basque Government (grant IT472-10) the Spanish MICINN
(Project No. FIS2009-12773-C02-02). MSB is supported by NSF 0545798.

\end{acknowledgements}

\end{document}